\documentclass[twocolumn,showpacs,prl,floatfix]{revtex4}
\usepackage{graphicx}
\usepackage{amsmath}
\def\beq{\begin{eqnarray}} \def\eeq{\end{eqnarray}}
\def\beqstar{\begin{eqnarray*}} \def\eeqstar{\end{eqnarray*}}
\newcommand{\bal}{\begin{align}} \def\eal{\end{align}}
\newcommand{\beqe}{\begin{equation}} \newcommand{\eeqe}{\end{equation}}
\newcommand{\p}[1]{(\ref{#1})}

\begin {document}
\title{Finite temperature effects in antiferromagnetism of  nuclear matter}
\author{ A. A. Isayev}
 \affiliation{Kharkov Institute of
Physics and Technology, Academicheskaya Str. 1,
 Kharkov, 61108, Ukraine
 }
 \date{\today}
\begin{abstract}   The influence of the finite temperature on the
antiferromagnetic (AFM) spin ordering in symmetric nuclear matter
with the effective Gogny interaction is studied within the
framework of a Fermi liquid formalism. It is shown that the AFM
spin polarization parameter of partially polarized nuclear matter
for low enough temperatures increases with temperature. The
entropy of the AFM spin state for some temperature range is larger
than the entropy of the normal state. Nerveless, the free energy
of the AFM spin state is always less than the free energy of the
normal state and, hence, the AFM spin polarized state is
preferable for all temperatures below the critical temperature.
\end{abstract}
 \maketitle

 \section{Introduction} The spontaneous appearance of  spin polarized states in nuclear
matter is a topic of a great current interest due to its relevance
in astrophysics. In particular, the effects of spin correlations
in the medium strongly influence the neutrino cross section and
neutrino mean free path. Therefore, depending on whether nuclear
matter is spin polarized or not, drastically different scenarios
of supernova explosion and cooling of neutron stars can be
realized. Another aspect relates to pulsars, which are considered
to be rapidly rotating neutron stars, surrounded by strong
magnetic field. One of the hypotheses to explain such a strong
magnetic field of a pulsar is that it can be produced by a
spontaneous ordering of spins in the dense stellar core.

If a spin polarized state appears, nucleons with spin up and spin
down lie on different Fermi surfaces. Spin-spin and spin-isospin
correlations, leading to the formation of a spin polarized state,
essentially depend on the overlap between Fermi surfaces. It is
controlled by the number densities of nucleons of different
species and spin polarization parameter, which, in turn, should be
determined  self-consistently.

From the general point of view, the problem of finding the phase
diagram of a many-particle Fermi system when fermions lie on
different Fermi surfaces is encountered in many physical cases.
For example, thermodynamic properties of a neutron-proton
condensate~\cite{ARS}--\cite{Is} in asymmetric nuclear matter are
governed by pairing correlations between neutrons and protons,
occupying two different Fermi spheres. Analogous situation appears
in  high density QCD, when the quark system is unstable against
the formation of the $\langle q\, q\rangle$ color superconducting
condensate~\cite{ARW}--\cite{PR}. Besides, one can mention the
spin singlet pairing in a superconducting metal in the presence of
magnetic impurities~\cite{LO,FF} or the B-phase of the superfluid
$^3$He (Balian-Werthamer phase) in a magnetic field~\cite{Le2},
when the spin degeneracy is loosened due to the Pauli
paramagnetism. Another example is the appearance of BCS pairing in
ultracold trapped quantum gases~\cite{RGJ,KHG,ZSS}.

 The
possibility of a phase transition of normal neutron and nuclear
matter to the ferromagnetic (FM) spin state was studied by many
authors~\cite{R}--\cite{BPM}, predicting the ferromagnetic
transition at $\varrho\approx(2$--$4)\varrho_0$ for different
parametrizations of Skyrme forces
($\varrho_0=0.16\,\mbox{fm}^{-3}$ is the nuclear matter saturation
density).  Competition between FM and  AFM spin ordering in
symmetric nuclear matter with the Skyrme effective interaction was
studied in Ref.~\cite{I}, where it was clarified that the FM spin
state is thermodynamically preferable to the AFM one for all
relevant densities. However,  strongly asymmetric nuclear matter
with Skyrme forces undergoes a phase transition to a state with
oppositely directed spins of neutrons and protons~\cite{IY}. The
same conclusion in favour of antiparallel ordering of neutron and
proton spins in symmetric nuclear matter was confirmed also in
Ref.~\cite{IY2} for the Gogny effective interaction, where it was
shown that the AFM spin state appears at
$\varrho\approx3.8\varrho_0$.

For the models with realistic nucleon-nucleon (NN) interaction,
the ferromagnetic phase transition seems to be suppressed up to
densities well above $\varrho_0$~\cite{PGS}--\cite{H}. In
particular, no evidence of ferromagnetic instability has been
found in recent studies of neutron matter~\cite{VPR} and
asymmetric nuclear matter~\cite{VB} within the
Brueckner--Hartree--Fock approximation with realistic Nijmegen II,
Reid93, and Nijmegen NSC97e NN interactions. The same conclusion
was obtained in Ref.~\cite{FSS}, where the magnetic susceptibility
of neutron matter was calculated with the use of the Argonne
$v_{18}$ two--body potential and Urbana IX three--body potential.

Here we continue the study of spin polarized states in nuclear
matter, using as a potential of NN interaction the effective
 Gogny forces~\cite{DG,BGG} and assuming that the AFM spin
 ordering is realized as a ground state of nuclear matter at zero
 temperature~\cite{IY2}.
The main emphasis will be laid on determining the finite
temperature behavior of the AFM spin polarization. The first goal
of the study is to show that at low enough temperatures thermal
fluctuations promote the AFM spin polarization of nuclear matter,
when a system of nucleons can be treated as a multicomponent Fermi
liquid~\cite{AKP,AIP,AIPY}. The second goal is to provide a fully
self-consistent calculation of the basic thermodynamic functions
of antiferromagnetically ordered nuclear matter at finite
temperatures with a modern effective finite range NN interaction.
In spite of that the entropy of the AFM spin state can be larger
than the entropy of the normal state, the total balance of free
energies lies with the AFM spin state for all temperatures below
the critical temperature.

\section{Basic Equations}

 The normal states of nuclear matter are described
  by the normal distribution function of nucleons $f_{\kappa_1\kappa_2}=\mbox{Tr}\,\varrho
  a^+_{\kappa_2}a_{\kappa_1}$, where
$\kappa\equiv({\bf{p}},\sigma,\tau)$, ${\bf p}$ is the momentum,
$\sigma(\tau)$ is the projection of spin (isospin) on the third
axis, and $\varrho$ is the density matrix of the system. Bearing
in mind  to consider the possibility of FM and AFM phase
transitions,  the normal distribution function $f$  and the
nucleon single particle energy $\varepsilon$
 can be expanded in the
Pauli matrices $\sigma_i$ and $\tau_k$ in spin and isospin
spaces
\begin{align} f({\bf p})&= f_{00}({\bf
p})\sigma_0\tau_0+f_{30}({\bf p})\sigma_3\tau_0\label{7.2}\\
&\quad + f_{03}({\bf p})\sigma_0\tau_3+f_{33}({\bf
p})\sigma_3\tau_3. \nonumber \\ 
\varepsilon({\bf p})&=
\varepsilon_{00}({\bf
p})\sigma_0\tau_0+\varepsilon_{30}({\bf p})\sigma_3\tau_0\label{7.3}\\
&\quad + \varepsilon_{03}({\bf
p})\sigma_0\tau_3+\varepsilon_{33}({\bf p})\sigma_3\tau_3.
\nonumber
\end{align}
Expressions for  the distribution functions
$f_{00},f_{30},f_{03},f_{33}$
 in
terms of the quantities $\varepsilon$ read~\cite{I,IY}
\begin{align}
f_{00}&=\frac{1}{4}\{n(\omega_{n\uparrow})+n(\omega_{p\uparrow})+n(\omega_{n\downarrow})
+n(\omega_{p\downarrow}) \},\nonumber
 \\
f_{30}&=\frac{1}{4}\{n(\omega_{n\uparrow})+n(\omega_{p\uparrow})-n(\omega_{n\downarrow})-
n(\omega_{p\downarrow})
\},\label{2.4}\\
f_{03}&=\frac{1}{4}\{n(\omega_{n\uparrow})-n(\omega_{p\uparrow})+n(\omega_{n\downarrow})-
n(\omega_{p\downarrow})
\},\nonumber\\
f_{33}&=\frac{1}{4}\{n(\omega_{n\uparrow})-n(\omega_{p\uparrow})-n(\omega_{n\downarrow})+
n(\omega_{p\downarrow})
\}.\nonumber
 \end{align} Here $n(\omega)=\{\exp(\omega/T)+1\}^{-1}$ and
\begin{gather}
\omega_{n\uparrow}=\xi_{00}+\xi_{30}+\xi_{03}+\xi_{33},\;\nonumber\\
\omega_{p\uparrow}=\xi_{00}+\xi_{30}-\xi_{03}-\xi_{33},\;\label{2.5}\\
\omega_{n\downarrow}=\xi_{00}-\xi_{30}+\xi_{03}-\xi_{33},\;\nonumber\\
\omega_{p\downarrow}=\xi_{00}-\xi_{30}-\xi_{03}+\xi_{33},\;\nonumber\end{gather}
where \begin{align*}\xi_{00}&=\varepsilon_{00}-\mu_{00},\;
\xi_{30}=\varepsilon_{30},\;
\\
\xi_{03}&=\varepsilon_{03}-\mu_{03},\;\xi_{33}=\varepsilon_{33},\\
\mu_{00}&={\frac{\mu_n+\mu_p}{2}},\quad
\mu_{03}={\frac{\mu_n-\mu_p}{2}}.\end{align*}
$\mu_n,\mu_p$ being the chemical potentials of neutrons and
protons. The branches $\omega_{n\uparrow},\omega_{n\downarrow}$ of
the quasiparticle spectrum correspond to neutrons with spin up and
spin down, and  the branches
$\omega_{p\uparrow},\omega_{p\downarrow}$ correspond to protons
with spin up and spin down.

The distribution functions $f$ should satisfy the normalization
conditions
\begin{align} \frac{4}{\cal
V}\sum_{\bf p}f_{00}({\bf p})&=\varrho,\label{3.1}\\
\frac{4}{\cal V}\sum_{\bf p}f_{03}({\bf
p})&=\varrho_n-\varrho_p\equiv\alpha\varrho,\label{3.3}\\
\frac{4}{\cal V}\sum_{\bf p}f_{30}({\bf
p})&=\varrho_\uparrow-\varrho_\downarrow\equiv\Delta\varrho_{\uparrow\uparrow},\label{3.2}\\
\frac{4}{\cal V}\sum_{\bf p}f_{33}({\bf
p})&=(\varrho_{n\uparrow}+\varrho_{p\downarrow})-
(\varrho_{n\downarrow}+\varrho_{p\uparrow})\equiv\Delta\varrho_{\uparrow\downarrow}.\label{3.4}
 \end{align}
 Here $\alpha$ is the isospin asymmetry parameter, $\varrho_{n\uparrow},\varrho_{n\downarrow}$ and
 $\varrho_{p\uparrow},\varrho_{p\downarrow}$ are the neutron and
 proton number densities with spin up and spin down,
 respectively;
 $\varrho_\uparrow=\varrho_{n\uparrow}+\varrho_{p\uparrow}$ and
$\varrho_\downarrow=\varrho_{n\downarrow}+\varrho_{p\downarrow}$
are the nucleon densities with spin up and spin down. The
quantities $\Delta\varrho_{\uparrow\uparrow}$ and
$\Delta\varrho_{\uparrow\downarrow}$ play the roles of  FM and AFM
spin order parameters~\cite{IY}.

The self--consistent equations for the components of the single
particle energy have the form~\cite{I,IY} \bal\xi_{00}({\bf
p})&=\varepsilon_{0}({\bf p})+\tilde\varepsilon_{00}({\bf
p})-\mu_{00},\;
\xi_{30}({\bf p})=\tilde\varepsilon_{30}({\bf p}),\label{14.2} \\
\xi_{03}({\bf p})&=\tilde\varepsilon_{03}({\bf p})-\mu_{03}, \;
\xi_{33}({\bf p})=\tilde\varepsilon_{33}({\bf
p}).\nonumber\end{align} Here $\varepsilon_0({\bf p})$ is the free
single particle spectrum,  and
$\tilde\varepsilon_{00},\tilde\varepsilon_{30},\tilde\varepsilon_{03},\tilde\varepsilon_{33}$
are the FL corrections to the free single particle spectrum,
related to the normal FL amplitudes $U_0({\bf k}),...,U_3({\bf k})
$ by formulas
\begin{align}\tilde\varepsilon_{00}({\bf
p})&=\frac{1}{2\cal V}\sum_{\bf q}U_0({\bf k})f_{00}({\bf
q}),\;{\bf k}=\frac{{\bf p}-{\bf q}}{2}, \label{14.1}\\
\tilde\varepsilon_{30}({\bf p})&=\frac{1}{2\cal V}\sum_{\bf
q}U_1({\bf k})f_{30}({\bf q}),\nonumber\\ 
\tilde\varepsilon_{03}({\bf p})&=\frac{1}{2\cal V}\sum_{\bf
q}U_2({\bf k})f_{03}({\bf q}), \nonumber\\
\tilde\varepsilon_{33}({\bf p})&=\frac{1}{2\cal V}\sum_{\bf
q}U_3({\bf k})f_{33}({\bf q}). \nonumber
\end{align}

   To obtain
 numerical results, we  use the  effective Gogny interaction D1S~\cite{BGG}. Expressions for
 the normal
 FL amplitudes in terms of Gogny force parameters were written in Ref.~\cite{IY2}.
 Thus,
with account of  expressions \p{2.4} for the distribution
functions $f$, we obtain the self--consistent equations \p{14.2},
\p{14.1}
 for the components of the single particle energy $\xi_{00}({\bf
p}),\xi_{30}({\bf p}),\xi_{03}({\bf p}),\xi_{33}({\bf p})$, which
should be solved jointly with the normalization conditions
\p{3.1}--\p{3.4}, determining the chemical potentials
$\mu_{00},\mu_{03}$,
 FM  and AFM spin
 order parameters
$\Delta\varrho_{\uparrow\uparrow}$,
$\Delta\varrho_{\uparrow\downarrow}$.

To examine the thermodynamic stability of different solutions of
self-consistent equations, it is necessary to compare the
corresponding free energies $F=E-TS$, where the energy functional
$E$ is characterized, in the general case, by four FL amplitudes
$U_0,...,U_3$ and the entropy reads \bal S&=-\sum_{\bf
p}\sum_{\tau=n,\,p}\,\sum_{\sigma=\uparrow,\,\downarrow}\{n(\omega_{\tau\sigma})\ln
n(\omega_{\tau\sigma})\nonumber\\ &\quad+\bar
n(\omega_{\tau\sigma})\ln \bar n(\omega_{\tau\sigma})\}, \;\bar
n(\omega)=1-n(\omega).\nonumber
\end{align}

\section{Phase transitions at finite temperature}
Further we will consider   symmetric nuclear matter
($\varrho_n=\varrho_p$). It was shown in Ref.~\cite{IY2} that in
symmetric nuclear matter with D1S Gogny interaction only the AFM
spin ordering is realized at zero temperature, but the
self--consistent equations have no solutions at all corresponding
to the FM spin ordering.
Our aim here is to study the temperature behavior of the AFM spin
polarization in the whole temperature domain below the critical
temperature, $T\leq T_c$.

 In
the AFM spin state of symmetric nuclear matter
$\varrho_{n\uparrow}=\varrho_{p\downarrow},
\varrho_{n\downarrow}=\varrho_{p\uparrow}$, neutrons with spin up
and protons with spin down   fill the Fermi surface of radius
$k_2$ and neutrons with spin down and protons with spin up occupy
the Fermi surface of radius $k_1$, satisfying at zero temperature
the equations \beqe
\frac{1}{3\pi^2}(k_2^3-k_1^3)=\Delta\varrho_{\uparrow\downarrow},\quad
\frac{1}{3\pi^2}(k_1^3+k_2^3)=\varrho.\nonumber\end{equation}

\begin{figure}[tb]
\includegraphics[height=7.0cm,width=8.6cm,trim=48mm 122mm 60mm 69mm,
draft=false,clip]{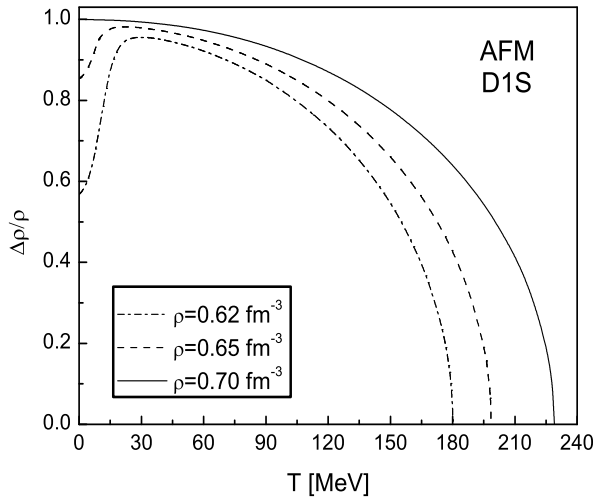}  \caption{ AFM spin polarization
parameter as a function of temperature at different densities for
the D1S
 Gogny force. }\label{fig1}
\end{figure}

Now we present the results of the numerical  solution of the
self--consistent equations with the  D1S  Gogny effective force.
In Fig.~1 it is shown the dependence of the AFM spin polarization
parameter $\Delta\varrho_{\uparrow\downarrow}/\varrho$ as a
function of temperature at different fixed densities. The
interesting feature is that, if at $T=0$ we have partially AFM
polarized state ($\Delta\varrho_{\uparrow\downarrow}/\varrho<1$),
then under increase of temperature within some temperature
interval the AFM spin polarization parameter increases as well.
This behavior is in contrast with the intuitive supposition  that
thermal fluctuations act as a destroying factor on  spin ordering.
Oppositely, for not too large temperatures thermal fluctuations in
nuclear matter promote the AFM spin ordering. The reason for such
a behavior is that thermal fluctuations smear Fermi surfaces of
nucleons, leading, thus, to increasing overlap between Fermi
surfaces. Since interaction between free nucleons is most strong
between a neutron and a proton in the spin triplet state, then,
mainly, due to this interaction, modified by the medium, some of
neutrons with spin down and  protons with spin up undergo spin
flip transitions from the inner Fermi surface to the outer one.
Thus, initial increase of AFM spin polarization with temperature
is a result of influence of thermal effects and medium
correlations. Under further increasing temperature thermal
fluctuations suppress AFM spin ordering, until it completely
vanishes.

\begin{figure}[t]
\includegraphics[height=7.0cm,width=8.6cm,trim=48mm 122mm 60mm 68mm,
draft=false,clip]{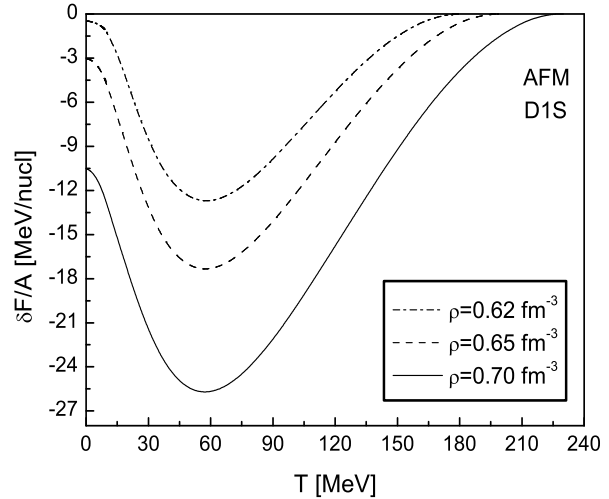}  \caption{ The free energy per
nucleon, measured from its value in the normal state, for the AFM
spin state  as a function of temperature at different densities
for the D1S  Gogny force.}
 \label{fig2}
\end{figure}

In Fig.~\ref{fig2},  the difference between the free energies per
nucleon  of the spin ordered and normal states is shown as a
function of temperature for different fixed densities. One can see
that, first, as a result of the initial increase of AFM spin
polarization, the free energy of the AFM spin state decreases with
temperature and after that the difference between the  free
energies of the AFM  and normal states becomes smaller, until it
vanishes at critical temperature, dependent on density.

\begin{figure}[tb]
\includegraphics[height=7.0cm,width=8.6cm,trim=48mm 122mm 60mm 69mm,
draft=false,clip]{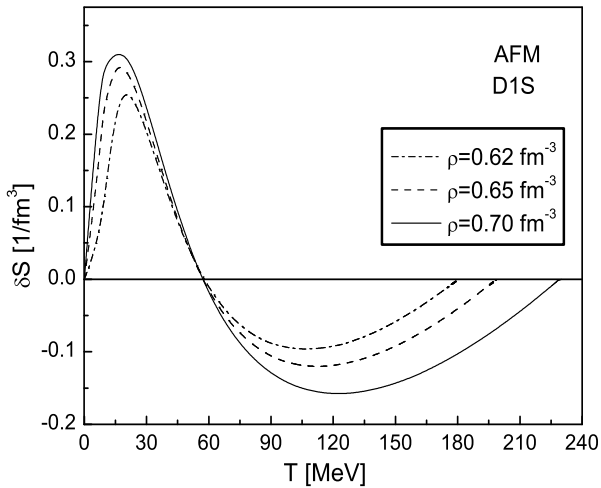}  \caption{ Same as in Fig.~2, but for
the density of entropy, measured from its value in the normal
state.}
 \label{fig3}
\end{figure}

Unexpected moment appears if we consider separately the
temperature behavior of the entropy of the AFM state. In
Fig.~\ref{fig3}, the difference between the densities of the
entropies of the AFM and normal states is shown as a function of
temperature. One can see that for low temperatures the entropy of
the AFM state  is larger than  the entropy of the normal state. It
looks like the AFM state at low finite temperatures is less
ordered than the normal state. Under further increasing
temperature the difference of the entropies changes its sign and
becomes negative, that corresponds to the intuitively expected
behavior. In spite of that the entropy of the AFM state may be
larger or less than the entropy of the normal state, the
difference of the free energies preserves its sign for all
temperatures  below $T_c$. Note that analogous anomalous behavior
of the entropy with temperature was observed also in superfluid
asymmetric nuclear matter~\cite{SL}. In that case for low enough
temperatures the entropy of the superfluid state is larger than
that for the normal state. Nerveless, as in our case, the total
balance of the free energies is preserved in favour of superfluid
state for all temperatures below the critical temperature. It is
worthy to note that for the AFM spin state anomalous behavior  of
the entropy is observed already in symmetric nuclear matter while
in the superfluid case it is observed only at finite isospin
asymmetry. The difference is that
 in the spin ordered state the
separation of Fermi surfaces is controlled by  the AFM spin
polarization parameter which, in turn, is not an independent
quantity, but should be found self-consistently.

Note that the stability of the equation of state of nuclear matter
with the Skyrme effective interaction in terms of Landau
parameters was examined in Ref.~\cite{MNG}, where the stability
conditions were formulated as the inequalities for the Skyrme
force parameters. This study is based on the approximation of the
effective mass, when the quadratic terms on momentum in the Skyrme
interaction are incorporated in the single particle spectrum. The
approximation of the effective mass, being independent of
temperature as in Ref.~\cite{MNG}, is a strong simplifying
assumption and cannot explain the change in the sign  of the
difference between entropies of polarized and unpolarized states
at certain temperature, as seen from Fig.~3. In the general case
of a finite range interaction, like Gogny force in our case or
Paris NN potential in Ref.~\cite{SL}, the single particle spectrum
is to be determined self-consistently by solving the corresponding
integral equations (in our case, Eqs.~\p{14.1}). Thus, the
anomalous behavior of the entropy with temperature should be
associated with the complicated renormalization of the free single
particle spectrum in a strongly interacting nucleon medium.

Note that, since in the present study only symmetric nuclear
matter has been  considered, the obtained results cannot be
directly extrapolated to proto-neutron stars, whose core
represents essentially asymmetric nuclear matter. For strongly
isospin asymmetric system, the appropriate choice of the Gogny
force is the use of the D1P parametrization~\cite{FES}, giving the
correct behavior of the energy per nucleon at high densities.

In summary, we have considered AFM polarized states in symmetric
nuclear matter with the effective Gogny interaction at finite
temperatures. It has been shown that  the AFM spin polarization
initially increases with temperature as a result of smearing Fermi
surfaces of nucleons due to thermal fluctuations and spin and
isospin dependent correlations in the medium. While the difference
of the entropies of the AFM and normal states anomalously changes
its sign at certain temperature, the total balance of the free
energies lies with the AFM spin state for all temperatures below
the critical temperature.



\begin{thebibliography}{99} 
\bibitem{ARS}  T. Alm, G. R\"opke, and M. Schmidt, Z. Phys. A
  {\bf 337}, 355 (1990).
\bibitem{VGD}  B.E. Vonderfecht, C.C. Gearhart, W.H. Dickhoff, A.
Polls, and A. Ramos,
  Phys.  Lett.   {\bf 253B}, 1 (1991).
\bibitem{BBL} M. Baldo, I. Bombaci, and U. Lombardo,
  Phys.  Lett.   {\bf 283B}, 8 (1992).
\bibitem{ARSW} T. Alm, G. R\"opke, A. Sedrakian, and F. Weber,  Nucl.
Phys.   {\bf A406}, 491 (1996).
\bibitem{SL}  A. Sedrakian, and U. Lombardo,
   Phys.  Rev. Lett.   {\bf 84}, 602 (2000).
\bibitem{Is}  A.A. Isayev,
   Phys.  Rev. C   {\bf 65}, 031302 (2002).
\bibitem{ARW}  M. Alford, K. Rajagopal, and F. Wilczek,
  Phys.  Lett.   {\bf 422B}, 247 (1998).
\bibitem{RSS}  R. Rapp, T. Sch\"afer, E.V. Shuryak, and M. Velkovsky,
   Phys.  Rev. Lett.   {\bf 81}, 53 (1998).
\bibitem{BR} D. Blaschke and C.D. Roberts,  Nucl.
Phys.   {\bf A642}, 197 (1998).
\bibitem{PR} R. Pisarski, and D. Rischke,  Phys. Rev. D {\bf 60}, 094013 (1999).
\bibitem{LO} A.I. Larkin, and Yu. N. Ovchinnikov,  Zh. Eksp. Teor. Fiz.
  {\bf 47}, 1136 (1964) [Sov. Phys. JETP {\bf 20}, 762 (1965)].
\bibitem{FF} P. Fulde and R. A. Ferrell,  Phys.  Rev.   {\bf 135}, 550
(1964).
\bibitem{Le2} A. Leggett,   Rev. Mod. Phys.  Rev.   {\bf 41}, 331
(1975).
\bibitem{RGJ}  C.A. Regal, M. Greiner, and D.S. Jin, Phys.  Rev. Lett.
 {\bf 92}, 040403 (2004).
\bibitem{KHG}  J. Kinast, S.L. Hemmer, M.E. Gehm, A. Turlapov, and J.E. Thomas,
 Phys.  Rev. Lett.
 {\bf 92}, 150402 (2004).
\bibitem{ZSS} M.W. Zwierlein, C.A. Stan, C.H. Schunck, S.M.F. Raupach, A.J. Kerman, and W. Ketterle,
 Phys.  Rev. Lett.
 {\bf 92}, 120403 (2004).
\bibitem{R}  M.J. Rice,   Phys.  Lett.   {\bf 29A}, 637 (1969).
\bibitem{S}  S.D. Silverstein,   Phys.  Rev. Lett.   {\bf 23}, 139 (1969).
\bibitem{O} E. {\O}stgaard,  Nucl.
Phys.   {\bf A154}, 202 (1970).
\bibitem{VNB} A.
Viduarre, J. Navarro, and J. Bernabeu,  Astron. Astrophys.
{\textbf {135}}, 361 (1984).
\bibitem{RPLP}  S. Reddy, M. Prakash, J.M. Lattimer, and J.A. Pons, Phys.  Rev.  C {\bf 59},
 2888 (1999).
\bibitem{ALP} A.I.  Akhiezer, N.V.  Laskin, and S.V.  Peletminsky,
Phys.  Lett.   {\bf 383B}, 444 (1996);  JETP {\bf 82}, 1066
(1996).
\bibitem{MNQN}  S. Marcos, R. Niembro, M.L. Quelle, and J. Navarro,   Phys.  Lett.   {\bf 271B},
277 (1991).
\bibitem{TT} T.  Maruyama and T. Tatsumi, Nucl.
Phys.   {\bf A693}, 710 (2001).
\bibitem{KW}  M. Kutschera, and W. Wojcik,   Phys.  Lett.   {\bf 223B}, 11 (1989).
\bibitem{BPM} A. Beraudo, A. De Pace, M. Martini, and A. Molinari,
Annals Phys. {\bf 311},  81 (2004); Arxiv: nucl-th/0409039.
\bibitem{I} A.A.  Isayev,    JETP Letters {\bf 77}, 251 (2003).
\bibitem{IY}  A.A.  Isayev, and  J. Yang,   Phys.  Rev.  C {\bf 69}, 025801 (2004).
\bibitem{IY2}  A.A.  Isayev, and  J. Yang,   Phys.  Rev.  C {\bf 70}, 064310 (2004).
\bibitem{PGS}  V.R. Pandharipande, V.K. Garde, and J.K. Srivastava,
   Phys.  Lett.   {\bf 38B}, 485 (1972).
\bibitem{BK}  S.O. B\"ackmann and C.G. K\"allman,
   Phys.  Lett.   {\bf 43B}, 263 (1973).
\bibitem{H}  P. Haensel, Phys.  Rev.  C {\bf 11},
 1822 (1975).
\bibitem{VPR}  I. Vida$\tilde{\mbox n}$a, A. Polls, and A. Ramos, Phys.  Rev.  C {\bf 65},
 035804 (2002).
\bibitem{VB}  I. Vida$\tilde{\mbox n}$a,  and I. Bombaci,  Phys.  Rev.  C {\bf 66},
 045801 (2002).
\bibitem{FSS}  S. Fantoni, A. Sarsa, and E. Schmidt,   Phys.  Rev. Lett.
 {\bf 87}, 181101 (2001).
 \bibitem{DG}  J.  Decharge and D. Gogny,   Phys.  Rev.  C {\bf 21}, 1568 (1980).
\bibitem{BGG}  J.F. Berger, M. Girod and D. Gogny,
   Comp. Phys. Comm. {\bf 63}, 365 (1991).
       \bibitem{AKP} A.I.
Akhiezer, V.V.  Krasil'nikov, S.V.  Peletminsky, and A.A.
Yatsenko, Phys. Rep.  {\bf 245}, 1 (1994).
\bibitem{AIP} A.I.  Akhiezer, A.A.  Isayev, S.V.  Peletminsky, A.P.
Rekalo, and A.A. Yatsenko,   JETP {\bf 85}, 1 (1997).
\bibitem{AIPY} A.I.  Akhiezer, A.A.  Isayev, S.V.  Peletminsky,
  and A.A. Yatsenko,   Phys.  Rev.  C {\bf 63}, 021304(R) (2001).
\bibitem{MNG} J. Margueron, J. Navarro, and N. V. Giai,   Phys.  Rev.  C {\bf 66}, 014303 (2002).
\bibitem{FES} M. Farine, D. Von-Eiff, P. Schuck, J.F. Berger, J. Decharge, and M. Girod,
   J. Phys. G {\bf 25}, 863 (1999).
\end{thebibliography}
\end{document}